# Higher-order-modes enhanced phase-matched dispersive-wave generation in the deep-blue and UV spectral region

X. T. Yang,[1,2,†] Z. Z. Luo,[1,2,†] J. P. Huang,[2,3,†] W. Y. Sun,[2] Y. Zheng[1,4], R.C. Yin[4], H. H. Yu,[1,*] M. Pang[2,3,*] and X. Jiang[1,3*]

*1 National Engineering Research Center of Fiber Optic Sensing Technology and Networks, Wuhan University of Technology, 122 Luoshi Road, Wuhan 430070, China*
*2 Key Laboratory of Materials for High Power Laser, Shanghai Institute of Optics and Fine Mechanics, Shanghai 201800, China*
*3 Russell Centre for Advanced Lightwave Science, Hangzhou Institute of Optics and Fine Mechanics, Hangzhou 311421, China*
*4 iFiber Optoelectronics Technology Co. Ltd., Ningbo 315100, China*

*Corresponding authors: hhyu@whut.edu.cn, pangmeng@siom.ac.cn, jiangx@whut.edu.cn;*

†These authors contributed equally.

**During the last few decades, solid-core photonic crystal fibers (PCFs) have been extensively explored to generate broadband, high-coherence supercontinua (SC). Limited by the material absorption and relatively low nonlinearity of fused silica, spectral broadening in silica PCF-based SCs is usually restricted to the blue to near-infrared spectral regions, even in developed commercial sources. The output spectra of these sources are missing short wavelengths of the full range. Many efforts have been spent to break the limitation. Among them, dispersive-wave (DW) generation has been investigated for triggering new frequencies in short wavelengths. With satisfied phase-matching conditions, excessive energy can be directly transferred from solitons of the anomalous dispersion region to DWs of the short wavelengths. However, a systematical study of factors, including phase-matched DWs, strongly related to the dispersion tailoring of higher-order modes (HOMs), has rarely been shown. This study reports the experimental observations of HOM-enhanced phase-matchings for the DW generation in the deep-blue and ultraviolet regions. A solid-core PCF-based, UV-extended SC source spanning a 2.8-octave-wide (350 nm to 2500 nm) is demonstrated. Meanwhile, we carefully verify our findings via numerical calculations. © 2022 Optical Society of America**

http://dx.doi.org/10.1364/OL.XXXXXX

Supercontinuum (SC) light sources have recently attracted lots of interest from academics to industry due to their great potential in astronomy, microscopy, spectroscopy, optical coherence tomography, biological imaging, etc. [1, 2]. The main challenge of an SC system lies in looking for optical materials with flexible dispersion tailoring and nonlinearity, to trigger multiple nonlinear processes for efficient light conversions [3]. Before the invention of photonic crystal fiber (PCF), highly nonlinear materials such as crystals, semiconductors, and dense liquids/gases were explored extensively as hosts for SC generation, normally with low efficiency and in bulk systems [3]. Solid-core PCF offers high freedom and feasibility in tailoring waveguide dispersion with enhanced nonlinearity to be an ideal platform for exciting many nonlinear optical phenomena [4]. It brings revolutionary changes, significantly enhancing the efficiency and stability of SC systems, and consequently, commercial products have become available. Unfortunately, on the one hand, due to unfavored dispersion profiles and relatively low nonlinear coefficient of fused silica, silica PCFs-based SCs have limited spectral extension in the 350-450 nm spectral region, even in commercial products such as LEUKOS DISCO-2 and NKT photonics FLU-6. On the other hand, broadband light sources covering the UV to the visible range are strongly requested for crucial applications in biomedical, molecular spectroscopy, and sensing areas [5-7]. Several techniques have been taken to realize SCs with UV extensions [8-12]. One is to use a 12 m long slowly-tapered PCF with a continuously-decreasing zero-dispersion-wavelength (ZDW) to achieve UV-enhanced SC down to 375 nm (10 dB) [8]. Another trial using fast tapered PCFs (5-30 mm taper lengths) extends the UV broadening to 280 nm [9]. Unfortunately, this system is unstable, and the UV light conversion can only last a few hours due to degradation. Using other fiber materials proves to be a successful alternative, when such a new glass fiber is also engineered with suitable dispersions in the UV region. For example, the following work reports an SC system with UV extension down to 200 nm, using a dispersion-tailored ZBLAN glass PCF [10]. Unfortunately, it requires additional equipment for novel glass and fiber fabrication, which is not an ordinary measure that general approaches can easily reach. Besides engineering the waveguide dispersion, another work using chirped pulses into a PCF shows UV generation at 370 nm [11]. Similarly, Wang et al. demonstrated a seven-core PCF, pumped by amplified chirped pulses, generating UV light conversion from 350 nm [12]. Dispersive wave (DW) generation, as a well-known nonlinear optical process, has also been investigated for UV-extended SC generation. A recent work reports the generation of DWs at 430 nm, with the help of higher-order-modes (HOMs) assisted dispersion tailoring [17]. Other nonlinear phenomena involving HOMs have also been extensively studied, including the four-wave mixing (FWM) [13], intermodal Raman scattering [14], third-harmonic generation (THG) [15], and SC generation [16, 17]. Nonlinear interactions using HOMs provide additional freedom for dispersion tailoring for light conversion based on different nonlinear phenomena.

This letter reports the experimental observations of HOMs-enhanced deep-blue and UV DW generations. An SC system is used as the platform for this study. The system consists of meter-long highly nonlinear PCFs, pumped by intense nanosecond pulses. Modulational instability (MI) and soliton dynamics dominate the initial spectral broadening from visible to mid-infrared. The further UV extension is mainly contributed by HOMs, with DW generation under satisfied phase-matching conditions. We systematically studied the procedure, both numerically and experimentally, using PCFs of various structures. Consequently, an optimized system demonstrates strong DW generation between 350 and 450 nm, in the forms of both the fundamental mode (FM) and HOMs.

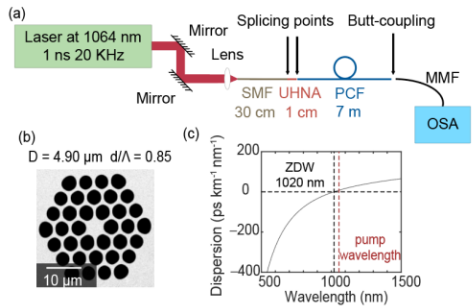

Fig. 1. (a) Experimental setup. SMF, single-mode fiber; PCF, photonic crystal fiber; UHNA, ultra-high numerical aperture. MMF, multimode fiber; OSA, optical spectrum analyzer. (b) Scanning electron microscopic image of the PCF. (c) Calculated wavelength-dependent dispersion of the PCF. ZDW, zero-dispersion wavelength.

Fig. 1(a) shows the experimental setup. It consists of the following components. A microchip laser delivering 1.1 ns pulses at 1064 nm is used as the pump laser. The repetition rate of this pulsed laser is 20 kHz, and the single-pulse energy is calculated at 15 µJ, when the maximum available average power of 300 mW is reached. To increase the stability and to reduce the insert losses between the pump and PCF, the output beam from the laser is firstly free-space coupled into a 30 cm long single-mode fiber (SMF), and then fusion spliced to a 1 cm ultra-high numerical aperture (UHNA) fiber, which is further spliced to a 7 m long highly nonlinear PCF. The UHNA fiber is used as a transition fiber to reduce splicing losses. The coupling efficiency from the light source to the SMF is ~80%, and the total splicing losses at two fusion points are effectively controlled under ~0.3 dB (transmission 93%) at 1064 nm, when splicing parameters are carefully optimized. At the output of the PCF, a multi-mode fiber is used for collecting output signals to an optical spectrum analyzer (OSA). Fig. 1(b) shows the scanning electron microscopic (SEM) photo of the PCF. The PCF has a ~4.9 µm solid silica core, surrounded by three rows of air channels in the cladding. The adjacent air-hole spacing (pitch Λ) is ~ 4.3 µm, and the air-filling fraction of the microstructured area is ~0.85. As shown in Fig. 1(c), the dispersion of the FM is checked by numerical calculation using the finite-element method (FEM). The ZDW of the FM locates at 1020 nm, so the pump wavelength at 1064 nm stays in the anomalous dispersion region.

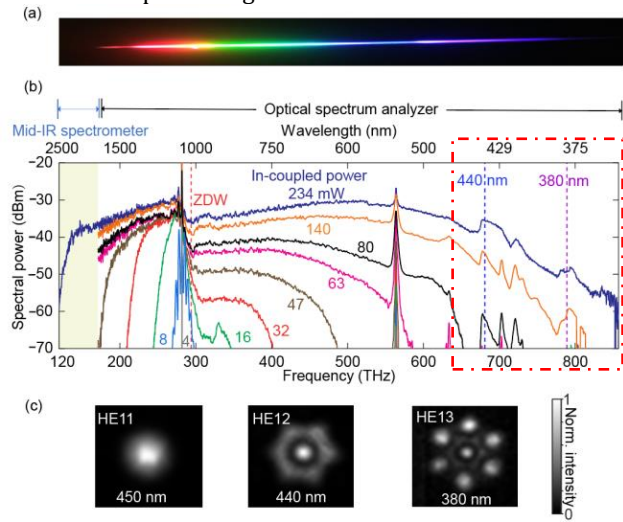

Fig. 2. (a) Dispersed output spectrum, cast onto a white screen using a fused-silica prism. (b) Spectral broadening at successively higher in-coupled powers up to ~ 234 mW. (c) Output far-field images at desired wavelengths analyzed via X-Y axis scanning using a 50-µm core multimode fiber as a probe (refer to the main text).

Fig. 2(b) shows the output spectra from the PCF, at successively higher in-coupled average power, starting from 4 mW. With the maximum available 234 mW average power coupled into the PCF, the system outputs a broad spectrum over-2.8-octave wide, spanning from 350 nm to 2500 nm. The light emitted from the PCF was dispersed by a fused-silica triple prism and cast onto a white screen, as shown in Fig. 2(a). Multiple spectral peaks in the UV to blue region, located between 380 nm (789 THz) and 440 nm (682 THz), are clearly seen in the dotted area of Fig. 2(b). We measured far-field transverse mode profiles via X-Y axial scanning using a 50 um-core multi-mode fiber as a probe. As seen in Fig 2(c), HE11, HE12, and HE13 modes, corresponding to the FM and two HOMs, are recognized at 450 nm, 440 nm, and 380 nm, respectively.

As seen in Fig. 2(b), even though the generated spectrum spans almost three-octave from the UV to the mid-IR, our interest remains on the short wavelength side. This is mainly due to three reasons: firstly, nonlinear light conversion in the UV-blue spectral region has proven to be a challenge [18]. For a long-time, this has been kept an open question in various SC systems using highly nonlinear PCFs [1, 2, 18]. Secondly, nonlinear DW generation in HOMs requires careful dispersion tailoring to satisfy phase-matchings, which brings lots of difficulties in fiber design and fabrication. Thirdly, most developed pump sources are in the near-IR, e.g., 0.8, 1.06, or 1.55 um. To generate light in short wavelengths, either we can use a visible/UV laser that may have unfavored specifications (low pulse energy; pump in the normal dispersion region, etc.) [19], or we must suffer from the low conversion efficiency to use a near-IR laser mentioned above. Because of the above reasons, silica PCFs-based SC sources usually miss short wavelengths of the full spectrum, even in the most developed commercial products such as LEUKOS DISCO-2 and NKT FLU-6. However, our system solves the problem that large portions of blue to UV light can be effectively generated (Fig. 2b), with the help of HOMs. It is therefore essential to understand the underlying mechanism.

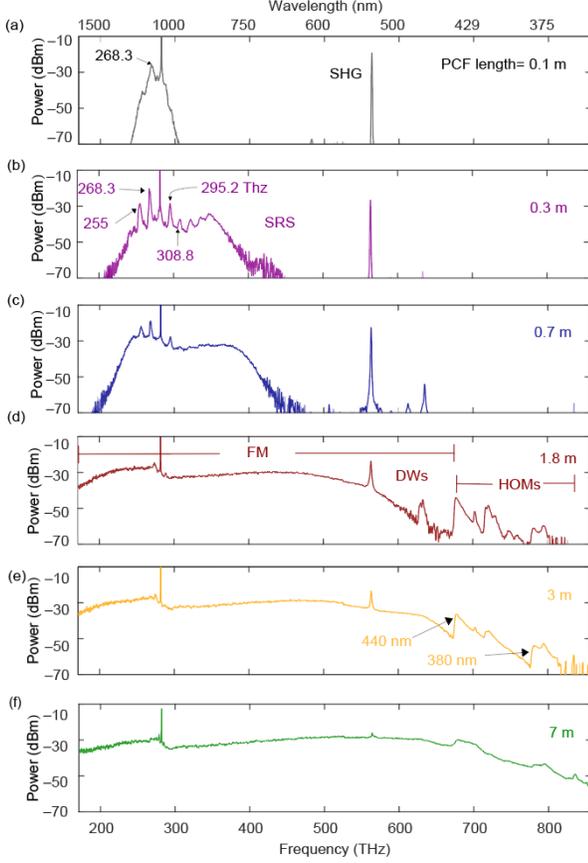

Fig. 3. Output spectra as a function of varied PCF lengths from 0.1 m to 7 m, at a fixed in-coupled power at 234 mW.

We verified our system by fixing the in-coupled power at 234 mW, then recorded the output spectra at varied PCF lengths from 0.1 m to 7 m (Fig. 3). The first observation is the second harmonic generation (SHG) at 532 nm, not generated from the PCF, but from the 30 cm, long germanium doped SMF (for mode-matching) core due to photon-induced color-center formation [20]. The critical feature of Fig. 3(a) is the presence of the spectral peak locating ~268.3 THz (1118 nm). The peak position is downshifted by ~13.2 THz with respect to the pump frequency (281.9 THz), corresponding to the maximum Raman gain of the fused silica [21]. The MI and cascaded SRS dominate the initial spectral broadening in the frequency domain (see Fig. 3(a), (b) and (c)). The pulse breaks into a trail of solitons in the time domain and the solitons undergo Raman-induced soliton self-frequency shift (SSFS), which accounts for the red-shift of spectrum in Fig. 3(d) [18]. Meanwhile, solitons with excessive energy emit their energy to DWs at the phase-matched wavelengths [1]. With the red-shift of the central wavelength of solitons, the phase-matched wavelengths gradually blueshift. However, due to the sharp change of the FM dispersion in the short wavelengths, the phase-matching conditions for DW generation in FM limit the blue-shift down to ~450 nm. Another striking feature of the spectra in Fig. (3)d is the presence of multi-spikes between 350 nm and 450 nm. The multi-spikes of deep-blue to UV are all in HOMs, based on our scanning test as we mentioned above. Considering the HOMs have lower effective refractive indices than FM, the FM solitons could transfer energy to DWs in the HOMs in the normal dispersion region, and the HOMs extend phase-matched wavelengths to deep-blue and UV regions, precisely making up for the missing sections (350 nm - 450 nm). With the help of HOMs, the spectrum can increase the blue shift of SC down to UV. The accumulation of nonlinear effects creates a flatter and smoother SC spectrum with the PCF length up to 7 m in Fig. 3(f).

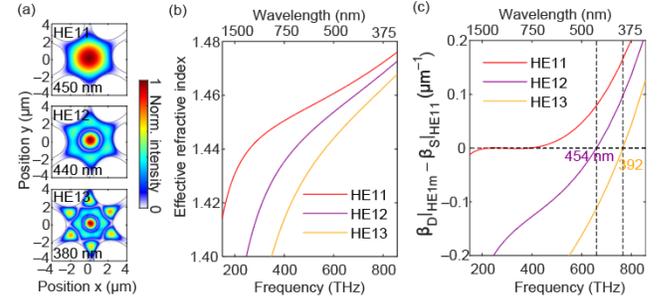

Fig. 4. (a) Calculated HE11, HE12 and HE13 mode profiles at 450 nm, 440 nm, and 380 nm. (b) Calculated frequency-dependent effective mode index of HE11, HE12, and HE13 mode. (c) Phase-matching curve for DW generation (see in the main text).

We also verified our experimental results via our numerical calculations. Fig. 4 shows the FEM results of the phase-matching conditions of HOM-enhanced DWs emission. The PCF structural model was vectorized from the SEM and imported to COMSOL Multiphysics software for the calculations. The HOMs, including HE11, HE12 and HE13 and wavelength dependent effective refractive indices, are calculated and shown in Fig. 4(a) and (b). Phase-matching conditions between solitons in FM and linear waves in HOMs requires $\beta_D|_{HE1m}(\omega) - \beta_S|_{HE11}(\omega) = 0$, where $\omega$ is the frequency and $\beta_D|_{HE1m}$ and $\beta_S|_{HE11}$ are the propagation constants of HE1m-mode linear waves and HE11-mode solitons. An analytical model takes the form [18]:

$$\beta_D|_{HE1m}(\omega) - \beta_S|_{HE11}(\omega) = \\ \beta_D|_{HE1m}(\omega) - \left[\beta_0 + \beta_1(\omega - \omega_0) + \frac{\gamma P_p}{2}\right],$$

Where $\beta_0$ and $\beta_1$ are the zero and first-order derivation of propagation constant of the FM, both evaluated at the soliton center frequency $\omega_0$, i.e., pump frequency, $\gamma$ is the nonlinear coefficient at $\omega_0$, $P_p$ is the peak power of the soliton and the nonlinear term $\gamma P_p/2$ makes a trivial contribution to the phase-matching. From Fig. 4(d) it is seen that the $\beta_D|_{HE1m}(\omega) - \beta_S|_{HE11}(\omega) = 0$ of HE12 mode and HE13 mode have a solution at 454 nm and 392 nm respectively, which slightly mismatches the experimental resonance peaks of HE12 mode ~440 nm and HE13 mode ~380 nm respectively. The mismatch between simulation and experimental results could be caused by simulation errors or the core inhomogeneity along PCF.

For achieving the controllable energy transfer, we have studied the influence of waveguide structure on phase-matching conditions.

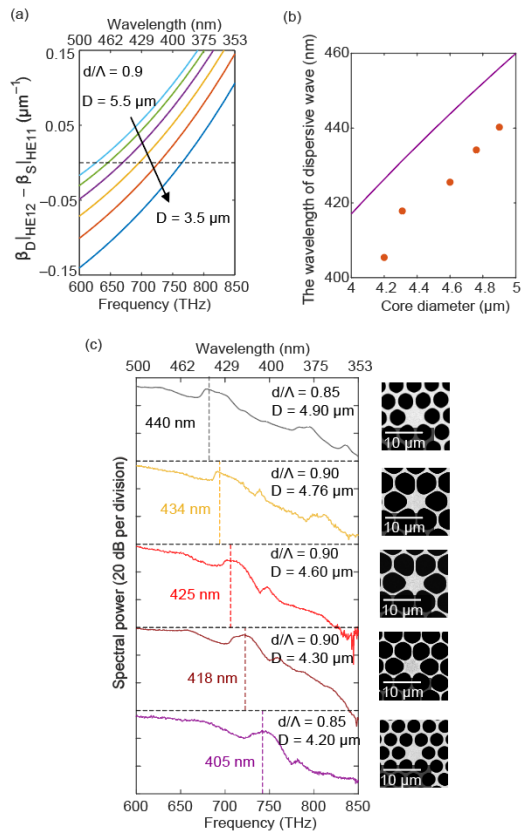

Fig. 5. (a) Calculated phase-matching curve for DW generation in HE12 mode using a perfect PCF model with the d/Λ 0.9 and different Ds. (b) The calculated core diameter-dependent wavelength of the DW. Orange dots are data points measured as shown in (c). (c) Experimental check using PCFs with different parameters and their SEM images.

We show the phase-matching curves for DWs for the HE12 modes using a perfect PCF model with a d/Λ of 0.9 and Ds from 5.5 μm to 3.5 μm in Fig. 5(a). The calculated wavelength of the DW as a function of D is presented in Fig. 5(b) (the orange dots were experimental results). For verifying our simulations, the SC spectrums of the other four PCFs with a d/Λ of ∼0.9 and Ds from ∼ 4.76 μm to ∼ 4.20 μm were tested. The tested results are shown in Fig. 5(c), corresponding to the orange dots in Fig. 5(b), and the images of SEM of the cross-section of PCFs are shown on the right panel. When the PCF core diameter decreases, the HE12 resonance peaks present a gradually blue-shifted trend from ∼ 440 nm to ∼ 405 nm, which is in good agreement with the theoretical prediction. Additionally, the lengths of the 4.90 μm-core PCF used in the experiment are ∼ 7 m and the other PCFs ∼ 5 m, which are long enough to generate the HOMs-enhanced peaks in the deep-blue and UV region.

In conclusion, we deeply discuss the process of the phase-matching condition of HOMs for extending the light conversion to the short wavelength region in detail. And we demonstrated that the solitons around pump wavelength transfer their energy to the DWs in the deep-blue and UV regions, in the form of HOMs, when the phase-matching conditions are satisfied, verified by using our numerical simulations. Secondly, we achieve the controllable HOMs-enhanced energy transfer of DWs to the short wavelength via tailoring the PCF waveguide structure. At the same time, the intensive nonlinear intermodal interactions extend the SC spectrum down to the UV region (350 nm), with many applications such as biomedical imaging and sensing.

**Funding.** Wuhan University of Technology.